\documentclass[reprint,twocolumn,superscriptaddress,nolongbibliography,aps,amsmath]{revtex4-2}
\usepackage{hyperref}
\usepackage{graphicx}
\usepackage{jabbrv}

\begin{document}

\title{Mid-infrared nonlinear optics in thin-film lithium niobate on sapphire}

\author{Jatadhari Mishra}
\thanks{These authors contributed equally to this work.\\Corresponding author: jmishra@stanford.edu}
\affiliation{E.\,L.\ Ginzton Laboratory, Stanford University, Stanford, California 94305, USA}

\author{Timothy P.\ McKenna}
\thanks{These authors contributed equally to this work.\\Corresponding author: jmishra@stanford.edu}
\affiliation{E.\,L.\ Ginzton Laboratory, Stanford University, Stanford, California 94305, USA}

\author{Edwin Ng}
\thanks{These authors contributed equally to this work.\\Corresponding author: jmishra@stanford.edu}
\affiliation{E.\,L.\ Ginzton Laboratory, Stanford University, Stanford, California 94305, USA}

\author{Hubert S.\ Stokowski}
\thanks{These authors contributed equally to this work.\\Corresponding author: jmishra@stanford.edu}
\affiliation{E.\,L.\ Ginzton Laboratory, Stanford University, Stanford, California 94305, USA}

\author{Marc Jankowski}
\affiliation{NTT Physics and Informatics Laboratories, NTT Research, Inc., East Palo Alto, CA 94303, USA}

\author{Carsten Langrock}
\affiliation{E.\,L.\ Ginzton Laboratory, Stanford University, Stanford, California 94305, USA}

\author{David Heydari}
\affiliation{E.\,L.\ Ginzton Laboratory, Stanford University, Stanford, California 94305, USA}

\author{Hideo Mabuchi}
\affiliation{E.\,L.\ Ginzton Laboratory, Stanford University, Stanford, California 94305, USA}

\author{M. M.\ Fejer}
\affiliation{E.\,L.\ Ginzton Laboratory, Stanford University, Stanford, California 94305, USA}

\author{Amir H.\ Safavi-Naeini}
\affiliation{E.\,L.\ Ginzton Laboratory, Stanford University, Stanford, California 94305, USA}

\begin{abstract}
Periodically poled thin-film lithium niobate (TFLN) waveguides have emerged as a leading platform for highly efficient frequency conversion in the near-infrared. However, the commonly used silica bottom-cladding results in high absorption loss at wavelengths beyond 2.5~{\textmu}m. In this work, we demonstrate efficient frequency conversion in a TFLN-on-sapphire platform, which features high transparency up to 4.5~{\textmu}m. In particular, we report generating mid-infrared light up to 3.66~{\textmu}m via difference-frequency generation of a fixed 1-{\textmu}m source and a tunable telecom source, with normalized efficiencies up to 200$\%$/W-cm$^2$.  These results show TFLN-on-sapphire to be a promising platform for integrated nonlinear nanophotonics in the mid-infrared.
\end{abstract}

\maketitle

In the past decade, novel mid-IR coherent light sources have developed rapidly, driven by interest in spectroscopic applications~\cite{Picque2019}. Semiconductor optoelectronic emitters, such as quantum- and interband-cascade lasers have become commercially available, while next-generation mid-IR quantum well emitters and quantum dot emitters seem promising~\cite{Yao2012, Jung2017}. Another class of mid-IR sources is based on frequency conversion in nonlinear optical media. A persistent effort in this area to produce devices with lower physical footprints and higher conversion efficiencies has enabled the steady progress of mid-IR sources from free-space systems~\cite{Burr1997} to diffused and diced large-core waveguides~\cite{Bchter2009, Mayer2016, Kowligy2018} to nanophotonic waveguides and microresonators in thin films~\cite{Singh2015, Gaeta2019}.

These latter nanophotonic implementations utilize the Kerr nonlinearity, which is relatively weak compared to the second-order nonlinearity accessible in competing thin-film platforms; efficient $\chi^{(2)}$-based frequency conversion in the near-IR has been demonstrated in a number of low-loss thin-film waveguide platforms such as lithium niobate (LN)~\cite{Wang18}, aluminium gallium arsenide~\cite{Chang2019}, and silicon nitride~\cite{Lu2020}. Among these, thin-film lithium niobate (TFLN) is the only material platform allowing for simple electric-field poling for quasi-phasematching (QPM)~\cite{Wang18} and integration of high-performance electro-optic (EO) devices~\cite{Wang2018}, as well as dispersion engineering~\cite{Jankowski2020}. As an example, broadband EO frequency-comb generation in the near-IR has been demonstrated in dispersion-engineered LN microring resonators, with octave-spanning EO combs appearing to be feasible~\cite{Zhang2019, Yu2019}. Such sources can be extended into the mid-IR, taking advantage of LN's transparency window up to 4.5~{\textmu}m~\cite{Myers1997, He2020}. However, the commonly used silica bottom-cladding layer suffers from high absorption loss at wavelengths beyond 2.5~{\textmu}m~\cite{Kitamura2007}.


In this work, we demonstrate, for the first time, efficient frequency conversion in periodically poled TFLN waveguides bonded to sapphire~\cite{Sarabalis2020}, which feature high transparency up to 4.5~{\textmu}m~\cite{Myers1997, Thomas1988}. We show CW mid-IR second-harmonic generation (SHG) of fundamental wavelengths spanning 2.75--3.27~{\textmu}m, and estimate normalized SHG conversion efficiencies up to 100$\%$/W-cm$^2$. Additionally, we demonstrate difference-frequency generation (DFG) between a 1-{\textmu}m pump and a tunable telecom source to produce CW mid-IR light at wavelengths spanning a range of 2.81--3.66~{\textmu}m. The normalized efficiency of this DFG process is approximately 200$\%$/W-cm$^2$. These devices have approximately two orders of magnitude higher normalized efficiencies compared to conventional periodically poled diffused LN waveguides~\cite{Bchter2009}.


A general, lossless CW three-wave interaction in a waveguide with a non-zero second-order susceptibility $\chi^{(2)}$ can be expressed in terms of the following coupled-mode equations:
\begin{eqnarray}
\partial_z A_{\omega_3}(z) &= -i\kappa_3 A_{\omega_2}(z)A_{\omega_1}(z) \exp(i\Delta \beta z),\label{eqn:TWM03}\\
\partial_z A_{\omega_2}(z) &= -i\kappa_2 A_{\omega_3}(z)A_{\omega_1}^*(z) \exp(-i\Delta \beta z),\label{eqn:TWM02}\\
\partial_z A_{\omega_1}(z) &= -i\kappa_1 A_{\omega_3}(z)A_{\omega_2}^*(z) \exp(-i\Delta \beta z),\label{eqn:TWM01}
\end{eqnarray}
where $\omega_3=\omega_2+\omega_1$ with $\omega_2 > \omega_1$, $A_{\omega_j}$ $(j \in \{1,2,3\})$ is the z-dependent amplitude of a waveguide mode at frequency $\omega_j$ with propagation constant $\beta_{\omega_j}$, normalized to have units of W$^{1/2}$, such that $|A_{\omega_j}|^2=P_{\omega_j}$ is the power contained in the field $\omega_j$. The phase-mismatch is given by $\Delta \beta = \beta_{\omega_3} - \beta_{\omega_2} - \beta_{\omega_1}$, and the nonlinear coupling is given by
\begin{equation}
\kappa_j = \frac{\sqrt{2 Z_0} \omega_j d_\mathrm{eff}}{c \sqrt{n_{\omega_{1}}n_{\omega_{2}}n_{\omega_{3}}A_{\mathrm{eff}}}},\label{eqn:kappaTWM}
\end{equation}
where $d_\mathrm{eff}$ is the effective nonlinear coefficient, and $A_\mathrm{eff}$ is the effective area of the nonlinear interaction in the waveguide~(\cite{Jankowski2020}, Supplement 1). 

In order to achieve phase-matching, we employ the technique of quasi-phasematching~\cite{Fejer1992}, where the ferroelectric domain orientation of the lithium niobate is periodically reversed with the periodicity $\Lambda_G = 2\pi/\Delta \beta$. For a first-order QPM grating with a $50\%$ duty cycle, and with all interacting waves polarized along the extraordinary axis, $d_\mathrm{eff}=2d_{33}/\pi$, where $d_{33}$ is the component of the susceptibility tensor along this axis and the largest tensor component in LN. We use $d_{33}$ = 20 pm/V for DFG and 19 pm/V for SHG in our simulations, based on Miller's Delta scaling. These values are known to approximately $\pm10\%$~\cite{Choy1976}, suggesting a $\pm20\%$ uncertainty in the simulated efficiency numbers.

In a phase-matched process, considering the low-conversion limit, the coupled-wave equations simplify to the following for the evolution of the field generated in a difference-frequency process:
\begin{equation}
\partial_z A_{\omega_1}(z) = -i\kappa_1 A_{\omega_3}(0)A_{\omega_2}^*(0), \label{eqn:TWM03_PM}
\end{equation}
from which we get the power scaling between input and output fields: $P_{\omega_1} = \eta_{0,\text{DFG}} P_{\omega_3} P_{\omega_2} L^2$, $L$ being the interaction length, and the normalized efficiency $\eta_{0,\text{DFG}}$ = ${\kappa_1}^2$, quoted in $\%$/W-cm$^2$. Experimentally, $\eta_{0,\text{DFG}}$ is determined by measuring the ratio $P_{\omega_1}/(P_{\omega_3} P_{\omega_2} L^2)$, taking these powers to be those inside the waveguide as inferred from the detected powers.

Similarly, for an SHG process, where $\omega_2 = \omega_1 = \omega$ and $\omega_3 = 2\omega$, the power scaling between fundamental and second harmonic in the low conversion limit is given by $P_{2\omega} = \eta_{0,\text{SHG}} P_{\omega}^2 L^2$, with $\eta_{0,\text{SHG}} = {\kappa_{2\omega}}^2/4$, and the experimental value of $\eta_{0,\text{SHG}}$ is determined by measuring $P_{2\omega}/(P_{\omega}^2 L^2)$.

We fabricated the device from a 4-inch wafer of magnesium-oxide-doped X-cut TFLN on a sapphire substrate with an initial film thickness $\sim630$~nm. To achieve quasi-phasematching in the 3-{\textmu}m band, the TFLN was periodically poled with periods in the range of 6.4--7.3~{\textmu}m using surface electrodes~\cite{Wang18}. Ridge waveguides were then fabricated in the poled TFLN using electron-beam lithography and argon-ion milling. Two different nominal top widths, 3.0~{\textmu}m and 3.3~{\textmu}m, were chosen for each poling period. The design etch depth was nominally 300~nm and measured to be 295~nm with an uncertainty of approximately 5~nm post fabrication. The side-wall angle of the fabricated waveguides was $\sim11.5$ degrees. In the final fabrication step, the device facets were prepared for end-fire coupling by laser dicing (DISCO DFL7341)~\cite{Jankowski2020}.
\begin{figure}[!htb]
\centering
\includegraphics[width=0.47\textwidth]{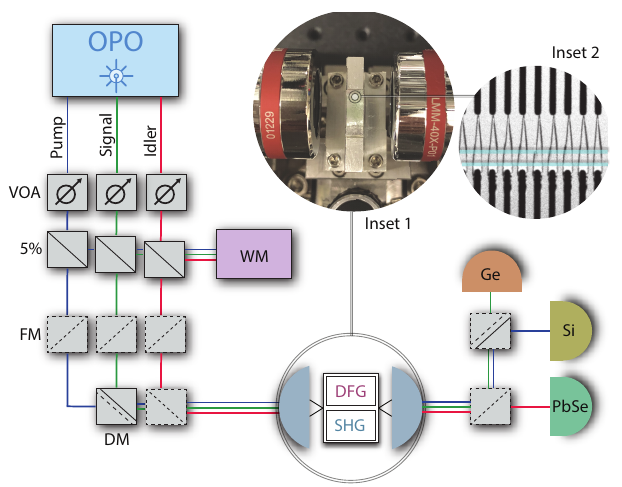}
\caption{Schematic of the experimental setup. The tunable OPO source outputs 1-{\textmu}m band (blue), 1.5-{\textmu}m band (green), and 3-{\textmu}m band (red) beams that are co-aligned in free space and incident on the TFLN waveguide chip via a reflective objective. The setup is switchable between SHG and DFG characterization. Inset 1 shows the chip in the end-firing characterization setup. Inset 2 shows periodic poling of TFLN between surface electrodes, taken via two-photon microscopy; the overlaid
lines show the approximate location of two adjacent waveguides. Abbreviations---VOA: variable optical attenuator, FM: flip mirrors, DM: dichroic mirrors, WM: wavelength meter.}
\label{fig:expt-setup}
\end{figure}

To characterize the TFLN waveguides, we use the output of a tunable CW OPO (Toptica TOPO), as shown in Fig.~\ref{fig:expt-setup}. Pumped by a 1.064~{\textmu}m amplified fiber laser, the OPO generates a tunable single-frequency signal (1.45--2.07~{\textmu}m) and an idler (2.19--4.00~{\textmu}m) in the $\sim$ 0.5--2 W power range. The third output is the residual pump. All three beam paths have variable attenuators for power control, followed by pick-off mirrors leading to a wavelength meter (Bristol 671). Before the waveguide, the three beams are co-aligned in free space via a series of flip and dichroic mirrors. The setup can be switched between the SHG experiment (idler input) and DFG experiment (pump and signal inputs). 

The waveguide setup uses reflective objectives (Thorlabs LMM-40X-P01) both to couple into the waveguide chip and to collect the output (Fig.~\ref{fig:expt-setup}, Inset 1). This enables all three input beams to be focused simultaneously at the input facet without incurring chromatic aberrations. We collect the mid-IR output on an amplified lead selenide photoconductive detector (Thorlabs PDA20H), the telecom-band output on a germanium power meter, and the 1-{\textmu}m output on a silicon power meter.

Knowledge of the absolute power of all interacting waves in the waveguide is required for efficiency measurements. We infer the power in the waveguide by measuring the power emitted from the output facet and dividing it by the collection efficiency of the objective, which was measured to be $\sim25\%\pm1\%$ at 1.064~{\textmu}m, $\sim33\%\pm1\%$ in the 1.5-{\textmu}m band, and $\sim27.5\%\pm2.5\%$ in the 3-{\textmu}m band. 

For measurements of the SHG normalized efficiencies, the idler of the OPO was set to the peak of the phase-matching curve of each device. We varied the idler beam power using a variable attenuator and observed the expected quadratic power of the detected SH. The fundamental and SH powers were measured at the output and their powers in the waveguide were estimated using the respective coupling efficiencies. A similar procedure was adopted for measurement of the DFG normalized efficiencies. However, instead of changing the input power which led to thermal drifts, the polarization of the input signal beam was rotated using a half-waveplate, while the power and polarization of the input pump beam were held constant. At the output, the pump was measured with a silicon power meter, and a polarizer was used to measure the TE component of the signal on a germanium power meter. The generated mid-IR was measured on the PbSe photoconductive detector. In this measurement, the DF power was observed to be linearly dependent on the signal power in the TE polarization. We estimate that the maximum number of generated mid-IR photons did not exceed $5\%$ of the number of photons in the weaker of the two inputs, which was the pump beam; this is commensurate with the low-conversion limit assumed in Eq.~\ref{eqn:TWM03_PM}. 




%

\begin{figure}[!htb]
\centering
\includegraphics[width=0.47\textwidth]{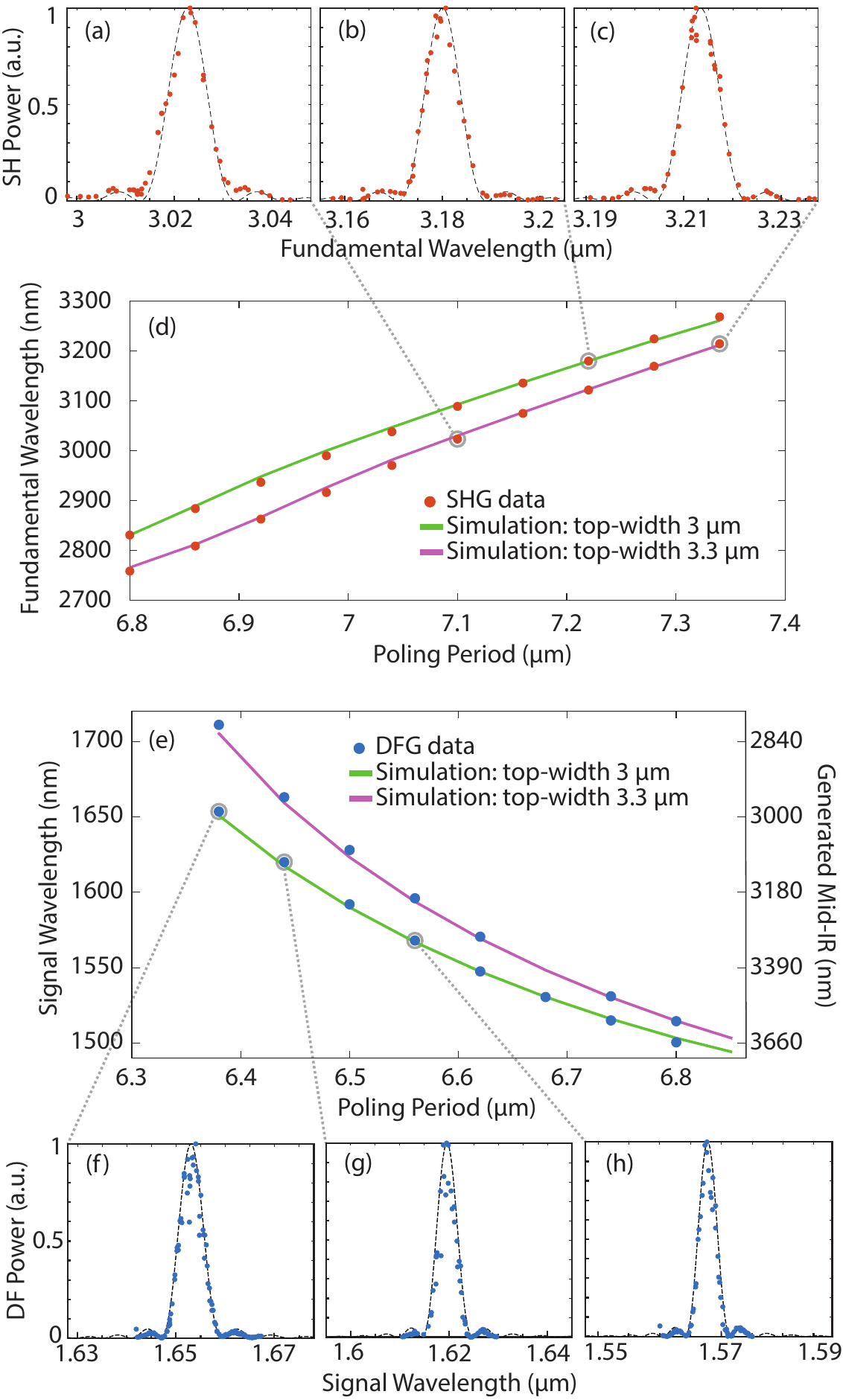}
\caption{(a--c) Measured SHG transfer functions (dots) vs.\ simulation (dashed lines). (d) SHG phase-matched fundamental wavelength as a function of poling period for two waveguide top-widths. (e) DFG phase-matching peak wavelength as a function of poling period for the two top-widths; input signal on left axis, generated mid-IR on right axis, for a 1.064-{\textmu}m pump. (f--h) Measured (dots) and simulated (dashed) DFG transfer functions for a fixed input pump at 1.064 {\textmu}m.}
\label{fig:TF-Tuning}
\end{figure}

In Fig.~\ref{fig:TF-Tuning}, we show that phase-matched SHG was observed in these nanophotonic devices with the idler from the OPO used as the fundamental spanning 2.75--3.27~{\textmu}m. For the two different waveguide top widths (3 and 3.3 {\textmu}m), we plot the fundamental wavelengths corresponding to the peaks of SHG phase-matching for ten different poling periods in Fig.~\ref{fig:TF-Tuning}(d) (dots). Figures~\ref{fig:TF-Tuning}(a-c) show three of the corresponding normalized SHG transfer functions (experimental data: dots, theory: dashed lines).

Also, in Fig.~\ref{fig:TF-Tuning}, we show that mid-IR DFG was observed spanning 2.81--3.66~{\textmu}m by mixing a tunable signal beam in the 1.5--1.72-{\textmu}m wavelength range against the fixed 1.064-{\textmu}m pump. For the two top widths, we plot the input signal wavelengths that show peak phase-matching for mid-IR generation for eight different poling periods in Fig.~\ref{fig:TF-Tuning}(e) (dots). Normalized DFG transfer functions for a fixed 1.064-{\textmu}m pump are shown in Figs.~\ref{fig:TF-Tuning}(f-h) for three devices. 

Based on an analysis of the measured phase-matched wavelengths compared to theoretical predictions, we took an offset of the phase-mismatch amounting to an effective poling period shift of $\sim$20~nm as a fitting parameter to account for small differences between the fabricated and simulated waveguide geometries. After taking this change into account, we observe for both SHG and DFG that the phase-matching tuning curves, with respect to poling periods, agree well with simulations (Figs.~\ref{fig:TF-Tuning}(d, e); solid lines). The shapes of the transfer functions and phase-matching bandwidths also closely match simulations for both processes, suggesting a sufficiently high fabrication fidelity over the entire $\sim$4.1-mm-long poled waveguide length. 

\begin{figure}
\centering
\includegraphics[width=0.47\textwidth]{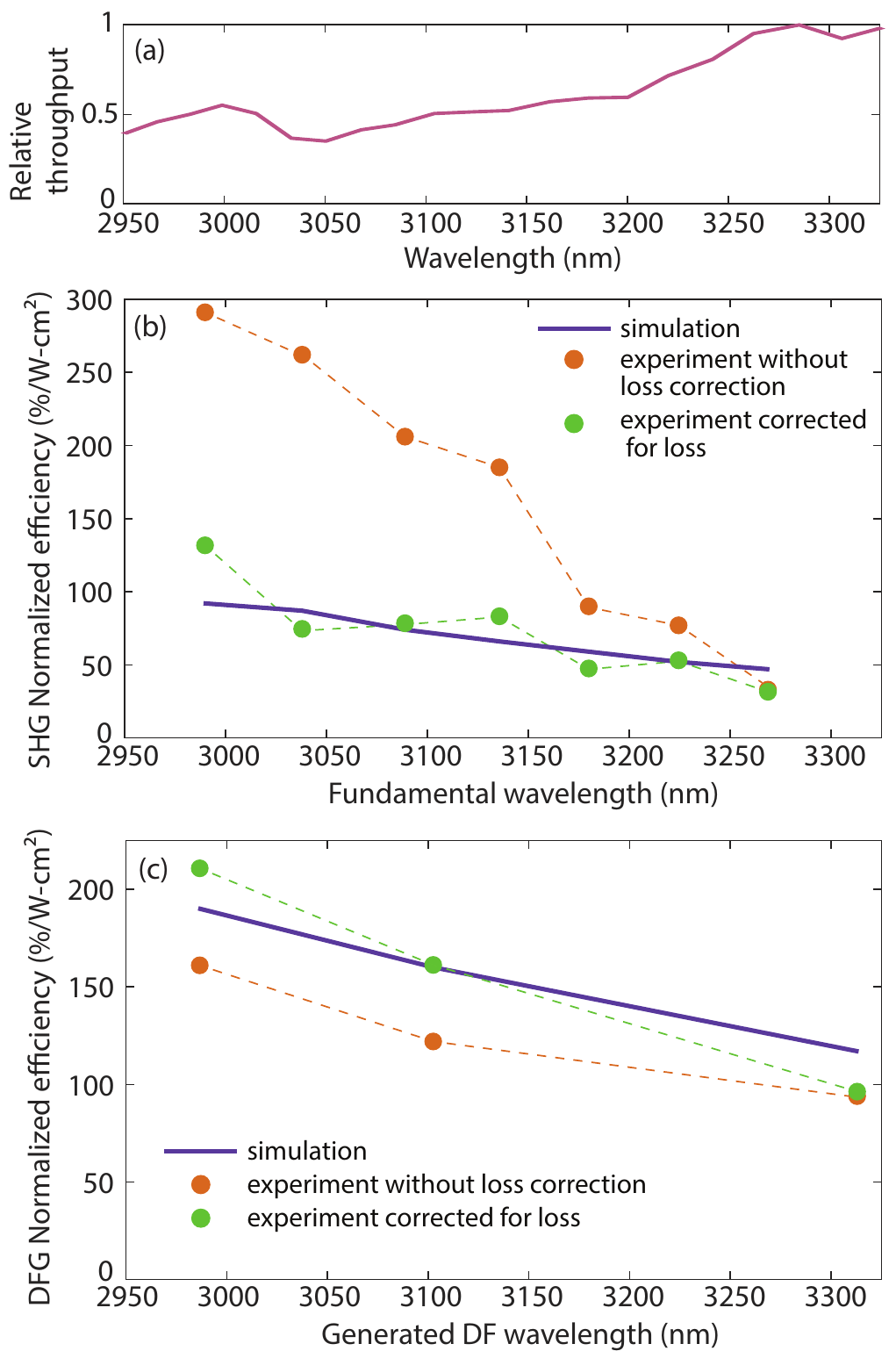}
\caption{(a) Normalized waveguide throughput in the mid-IR. (b) Experimentally measured normalized SHG efficiencies as computed from the measured output powers from the waveguide, and with corrections for the measured wavelength-dependent attenuation based on the mid-IR throughput, plotted alongside the simulated SHG efficiencies. (c) Experimentally measured normalized DFG efficiencies measured both with and without corrections for the wavelength-dependent attenuation of the long-wavelength, plotted alongside the simulated values.}
\label{fig:Norm-Eff}
\end{figure}

The simulated and measured normalized efficiencies for both the second-harmonic and difference-frequency process of mid-IR light are shown in Fig.~\ref{fig:Norm-Eff}. Due to their tight mode confinement, these TFLN devices feature predicted efficiencies for both processes higher by approximately two orders of magnitude than achievable in diffused waveguide platforms~\cite{Bchter2009}.

Figure~\ref{fig:Norm-Eff}(b) shows the experimentally measured normalized SHG efficiencies (dots) from seven different waveguides, alongside their simulated values (solid line) as a function of fundamental wavelength. However, the inferred efficiencies were significantly higher than predicted, with deviations getting larger towards shorter wavelengths. These initial measurements did not take the wavelength-dependent throughput into account, which was found to be a strong function of frequency in the 3-{\textmu}m band. This frequency-dependent throughput, shown in Fig.~\ref{fig:Norm-Eff}(a), was observed to be independent of the choice of waveguide and coupling optimization. We believe the origin of this loss is absorption due to OH either adsorbed on the surface or in the bulk LN material; further investigations of the mechanism are in progress. We note that annealing methods to drive hydrogen from LN crystals are available~\cite{Schwesyg2010}. We calculated the exponential loss coefficient that would contribute to the measured power loss over the length of the waveguide, and used this coefficient (varying between 1.7 and 0 cm$^{-1}$ over the range 2.95 to 3.325 {\textmu}m) to correct the measured fundamental power at the output. These loss-corrected efficiencies (Fig.~\ref{fig:Norm-Eff}(b)) are in good agreement with the simulated efficiencies. The measured transfer functions (Fig.~\ref{fig:TF-Tuning}) are also consistent with simulations that take into account the loss in the long-wave. 

In Fig.~\ref{fig:Norm-Eff}(c), we similarly plot the experimentally measured (dots) normalized DFG efficiencies from three different waveguides, both with and without loss correction, alongside the simulated efficiency values (solid line), as a function of the DFG wavelength. It is clear from these plots that the excess loss in mid-IR affects the estimate of the DFG efficiency linearly, while the SHG efficiency is affected quadratically. We note that the measured SHG and DFG normalized efficiencies can only be considered accurate within a few tens of percent, given the uncertainties in the calibration of internal waveguide power vs. detected output power.

In conclusion, we have demonstrated that periodically poled TFLN on sapphire is an efficient platform for second-harmonic and difference-frequency generation of mid-IR light due to tight optical confinement, wide transparency range, and large second-order nonlinearity, resulting in normalized efficiencies around two orders of magnitude higher than in diffused lithium niobate waveguides. Dispersion engineering along with efficient integration of electro-optic components in this platform could allow broadband comb-generation in the mid-IR, resulting in a useful source for spectroscopic applications. The highly efficient three-wave mixing seen here would also allow for effective frequency up-conversion of mid-IR light to the 1-um band for detection with silicon detectors, or down-conversion of telecom signals to mid-IR for robust  free-space optical data transmission, particularly in the atmospheric transmission window around 3.8 {\textmu}m~\cite{Bchter2009}.

\section*{Funding}
The authors wish to thank NTT Research for their financial and technical support through award 146395. This work was also supported by the NSF (CCF-1918549, PHY-2011363), the DARPA Young Faculty Award (DARPA-RA-18-02-YFA-ES-578), and the Department of Energy (DE-AC02-76SF00515). Devices were fabricated at the Stanford Nanofabrication Facility (NSF ECCS-1542152), the Stanford Nano Shared Facilities (NSF ECCS-2026822), and the Cell Sciences Imaging Facility (NCRR S10RR02557401). H.S. acknowledges support from the Urbanek Family Fellowship.

\section*{Disclosures}
The authors declare no conflicts of interest.

\bibliographystyle{osajnlnt}
\bibliography{references}

\end{document}